\documentclass[12pt,a4paper]{article}
\usepackage{amsmath}
\input epsf
\usepackage[dvips]{graphicx}
\usepackage{times}

\begin{document}
\begin{titlepage}
\title{Comment on confinement and unitarity  interrelation}
\author{ S.M. Troshin,
  N.E. Tyurin\\[1ex] \small\it Institute
\small\it for High Energy Physics,\\\small\it Protvino, Moscow Region, 142281 Russia}
\date{}
\maketitle
\begin{abstract}
We discuss  how confinement could result in the rational unitarization scheme. 
\end{abstract}
\vfill
\end{titlepage}

One of the fundamental
problems of  QCD  is related to confinement  of color.
On the other hand the hypothesis 
on the completeness of the set of asymptotic hadronic states plays an important role  (cf. \cite{bart}) in strong interaction dynamics
and leads, e.g. due to  unitarity of scattering matrix to the optical theorem relating the total cross-section
with  the forward elastic scattering amplitude.
Unitarity is formulated
for the asymptotic colorless  on-mass shell states and is not directly connected
to the fundamental fields of QCD -- colored fields of quarks and gluons. The  point of completeness can be
considered as a questionable in the QCD era. It might be reasonable to claim that set of hadronic states
do not provide complete set of states and unitarity in the sense indicated above could be violated (cf. \cite{rlst}).
It was stated in the indicated paper that the Hilbert space which corresponds to colorless hadron states and is constructed using vectors  spanned
on the physical vacuum should be extended. 
Indeed, nowadays it is  accepted that this vacuum state  is not
unique and colored  current quarks and gluons
are the degrees of freedom related to the perturbative vacuum.
According to the confinement property
of QCD, isolated colored objects cannot exist in the physical vacuum. Transition from physical vacuum 
to the perturbative one occurs in the  process of deconfinement and results in quark-gluon plasma
formation, i.e.   gaseous state of free colored quarks and gluons.   It is clear that
hadrons  and free quarks and gluons  cannot coexist  together since
they  live in different vacua \cite{mingmei} and  one can suppose that second Hilbert space should be
introduced\footnote{We do not concern here the possible existence of the third vacuum state  and related problems\cite{deconf}.}.

We address  above questions using as a guide the paper by N.N. Bogolyubov \cite{nnb}\footnote{The reference 
to this paper
was brought to our attention by V.A. Petrov.} and consider a state vector $\Phi$ being a sum of the two vectors
\[
 \Phi=\varphi+\omega.
\]
where $\varphi$ corresponds to physical states and belongs to the Hilbert space ${\cal H}_\varphi$ and $\omega$ -- to confined states and
belongs to the Hilbert space ${\cal H}_\omega$.  So, we  have that $\varphi={\cal P}\Phi$ and $\omega=(1-{\cal P})\Phi$, 
where $\cal P$ is a relevant projection operator.
 The difference with consideration performed in \cite{nnb} is in the replacement of the
states with indefinite metrics  to the  states of confined objects such as quarks and gluons.
The total Hilbert space $\cal H$ is a sum:
\[
\cal H={\cal H}_\varphi+{\cal H}_\omega
\]
and scattering operator $\cal \tilde S$ ($\Phi_{out}={\cal \tilde S}\Phi_{in}$) operating in $\cal H$ 
should not, in general, conserve probability and obey unitarity
since $\cal H$ includes ${\cal H}_\omega$ -- Hilbert space where confined objects live. Next, let impose condition similar to the one used in \cite{nnb}
\[
 \omega_{in}+\omega_{out}=0.
\]
 It means that {\it in-} and {\it out-} vectors 
corresponding to the states of the confined objects are just the mirror reflections of each other. 
Then the rational form of unitary scattering operator $\cal S$ 
\[
 {\cal S}=(1-{\cal U})(1+{\cal U})^{-1},
\]
in the physical Hilbert space ${\cal H}_\varphi$ ($\varphi_{out}={\cal S}\varphi_{in}$)
can easily be obtained, since
\[
\varphi_{out}={\cal P}{\cal \tilde S}( \varphi_{in}+ \omega_{in}).
\]
The operator $\cal U$ has the following form 
\[
{\cal U}= (1-{\cal P}){\cal \tilde S}.
\]
Thus, proceeding that way, we started with non-unitary scattering operator ${\cal \tilde S}$ and arrived
to the scattering operator ${\cal  S}$, which automatically satisfy unitarity condition.
Crucial assumption there was constraint for the states of confined objects $ \omega_{in}+\omega_{out}=0$, which
we supposed to be  a cofinement condition. Thus, it is very tempting to claim  that unitarity is related to confinement.
However, this claim should be taken with caution due to not completely clear status of the above 
constraint for the {\it in-} and {\it out-} vectors of states of confined
objects.

Rational or $U$-matrix form of unitarization was proposed long time ago \cite{heit} in the theory of radiation dumping,
self-damping of inelastic channels was considered in \cite{bbla} and
 for relatvistic case it was obtained in \cite{umat}. But, importance of the forgotten paper \cite{nnb} is in fact that
it provides clue for the physical interpretation of $U$-matrix. Nowadays it can be used for construction of a bridge
between physical states of hadrons and states of confined objects -- quarks and gluons. 

The elastic scattering $S$-matrix (i.e. the $2\to 2$  matrix element of the operator $\cal S$)
in the impact parameter representation can be
written (in this  unitarization scheme) in the form of linear 
rational transform (cf. \cite{inta} and references therein) and in
the case of pure imaginary $U$-matrix
\begin{equation}
S(s,b)=\frac{1-U(s,b)}{1+U(s,b)}, \label{um}
\end{equation}
where $U(s,b)$ is the generalized reaction matrix. 

It is
considered to be an input dynamical quantity. And this is essential point, the explicit form of the function $U(s,b)$ and numerical
predictions for the observable quantities depend on the particular  model used for hadron scattering description.
With account of what was said above we can associate this function with matrix elements\footnote{Imaginary part of $U(s,b)$ 
gets contributions from inelastic intermidiate channels.}
 of the operator
$(1-{\cal P}){\cal \tilde S}$, i.e. $U(s,b)$ should be related to the scattering dynamics of confined hadron constituents.
Rational representation of scattering matrix leads to several distinctive features such as dominating behavior of elastic scattering
cross-section, i.e. reflective scattering, peripherality of inelastic processes \cite{inta} and restoration of confined phase of hadronic
matter at very high temperatures \cite{deconf}. 
\section*{Acknowledgement}
We gratefully acknowledge stimulating discussions (and correspondence) with V.A. Petrov and J.P. Ralston.

\end{document}